\documentstyle[prl,aps]{revtex}

\begin{document}
\draft
\title{Spherical Scalar Field Halo in Galaxies}
\author{Tonatiuh Matos,$^{1}$
F. Siddhartha Guzm\'{a}n$^{1}$\thanks{
E-mail: siddh@fis.cinvestav.mx} and Dar\'{\i}o N\'{u}\~{n}ez$^{2}$}
\address{$^{1}$Departamento de F\'{\i}sica,
Centro de Investigaci\'on y de Estudios Avanzados del IPN,
A.P. 14-740, 07000 M\'exico D.F., MEXICO.\\
}
\address{$^{2}$Instituto de Ciencias Nucleares, 
Universidad Nacional Aut\'onoma de M\'exico,
A.P. 70-543, 04510 M\'exico D.F., MEXICO}
\date{\today}
\maketitle

\begin{abstract}
We study a spherically symmetric fluctuation of scalar dark matter in the 
cosmos and show that it could be the dark matter in galaxies, provided
that the scalar field has an exponential potential whose overall sign is
negative and whose exponent is constrained observationally by the rotation
velocities of galaxies. The local space-time of the fluctuation contains a
three dimensional space-like hypersurface with surplus of angle.
\end{abstract}



The existence of dark matter in the Universe has been firmly established by
astronomical observations at very different length-scales, ranging from
single galaxies, to clusters of galaxies, up to cosmological scale (see for
example \cite{falta}). A large fraction of the mass needed to produce the
observed dynamical effects in all these very different systems is not seen.
At the galactic scale, the problem is clearly posed: The measurements of
rotation curves (tangential velocities of objects) in spiral galaxies show
that the coplanar orbital motion of gas in the outer parts of these galaxies
keeps a more or less constant velocity up to several luminous radii \cite
{persic}, forming a radii independent curve in the outer parts of the
rotational curves profile; a motion which does not correspond to the one
due to the observed matter distribution, hence there must be present some
type of dark matter causing the observed motion. The flat profile of the
rotational curves is maybe the main feature observed in many galaxies. It is
believed that the dark matter in galaxies has an almost spherical
distribution which decays like $1/r^{2}$. With this distribution of some
kind of matter it is possible to fit the rotational curves of galaxies quite
well \cite{begeman}. Nevertheless, the main question of the dark matter
problem remains; which is the nature of the dark matter in galaxies? The
problem is not easy to solve, it is not sufficient to find out an exotic
particle which could exist in galaxies in the low energy regime of some
theory. It is necessary to show as well, that this particle (baryonic or
exotic) distributes in a very similar manner in all these galaxies, and
finally, to give some reason for its existence in galaxies.\newline

In previous works it has been explored, with considerable success, the
possibility that scalar fields could be the dark matter in spiral galaxies
by assuming that the scalar dark matter distributes as an axially symmetric
halo \cite{guzman1,guzman2}. The idea of these works is to explore whether
a scalar
field can fluctuate along the history of the Universe and thus forming
concentrations of scalar field density. If, for example, the scalar field
evolves with a scalar field potential $V(\Phi )\sim \Phi ^{2}$, the
evolution of this scalar field will be similar to the evolution of a perfect
fluid with equation of state $p=0$, $i.e.$, it would evolve as cold dark
matter \cite{turner}. However, it is not clear whether a spherical scalar
field fluctuation can serve as dark matter in galaxies. In this letter we
show that this could be the case. We assume that the halo of a galaxy is a
spherical fluctuation of cosmological scalar dark matter and study the
consequences for the space-time background at this scale, in order to
restrict the state equation corresponding to the dark matter inside the
fluctuation. We start from the general spherically symmetric line element 
and find
out the conditions on the metric in order that the test particles in the
galaxy possess a flat rotation curve in the region where the scalar field
(the dark matter) dominates. Finally we show that a spherical fluctuation of
the scalar field could be the dark matter in galaxies.\newline

Assuming thus that the dark matter is scalar, we start with the energy
momentum tensor $T_{\mu \nu }=\Phi _{,\mu }\Phi _{,\nu }-1/2g_{\mu \nu }\Phi
^{,\sigma }\Phi _{,\sigma }-g_{\mu \nu }V(\Phi )$, being $\Phi $ the scalar
field and $V(\Phi )$ the scalar potential. The Klein-Gordon and Einstein
equations respectively are:

\begin{eqnarray*}
&&\Phi _{;\mu }^{;\mu }-\frac{dV}{d\Phi }=0 \\
R_{\mu \nu } &=&\kappa _{0}[\Phi _{,\mu }\Phi _{,\nu }+g_{\mu \nu }V(\Phi )],
\end{eqnarray*}

\noindent where $R_{\mu \nu }$ is the Ricci tensor, $\sqrt{-g}$ the
determinant of the metric, $\kappa _{0}=8\pi G$ and a semicolon stands for
covariant derivative according to the background space-time; $\mu ,\nu
=0,1,2,3$. \newline

Assuming that the halo has spherical symmetry and that dragging effects on
stars and dust are inappreciable, i.e. the space-time is static, the
following line element is the appropriate

\begin{equation}
ds^{2}=-B(r)dt^{2}+A(r)dr^{2}+r^{2}d\theta ^{2}+r^{2}\sin ^{2}\theta
d\varphi ^{2}  \label{metric}
\end{equation}

\noindent where $A$ and $B$ are arbitrary functions of the coordinate $r$.
Following the analysis made for axisymmetric stationary space-times \cite
{nuestro}, we consider the Lagrangian for a test particle travelling on the
space time described by (\ref{metric}) which is

\begin{equation}
2{\cal {L}}=-B\dot{t}^{2}+A\dot{r}^{2}+r^{2}\dot{\theta}^{2}+r^{2}\sin
^{2}\theta \dot{\varphi}^{2}  \label{lagrangian}
\end{equation}

\noindent where a dot means derivative with respect to the proper time. From
(\ref{lagrangian}) the generalized momenta read:

\begin{eqnarray}
p_{t} &=&-E=-B\dot{t}  \label{pt} \\
p_{r} &=&A\dot{r}  \label{pr} \\
p_{\theta } &=&L_{\theta }=r^{2}\dot{\theta}  \label{pthe} \\
p_{\varphi } &=&L_{\varphi }=r^{2}\sin ^{2}\theta \dot{\varphi}  \label{pp}
\end{eqnarray}

\noindent being $E$ the total energy of a test particle and $L_{i}$ the
component of its angular momentum. It can be defined the Hamiltonian ${\cal H
}=p^{\mu }\dot{q_{\mu }}-{\cal L}$ and after rescalling the proper time for
the lagrangian to equal $1/2$ for time-like geodesics, the geodesic equation
for material particles (stars and dust) arises

\begin{equation}
\dot{r}^{2} + \frac{1}{A}[1 + \frac{L_{T}^{2}}{r^{2}} -
\frac{E^{2}}{B}]=0
\label{geodesic}
\end{equation}


\noindent being $L_{T}^{2} = L_{\theta }^{2}+\frac{L_{\varphi
}^{2}}{\sin ^{2}\theta }$ the first integral corresponding to the squared
total angular momentum. We are interested in circular and stable motion of
test particles, therefore the following conditions must be 
satisfied\newline

i) $\dot{r}=0$, circular trajectories\newline

ii)${\frac{{\partial V(r)}}{{\partial r }}}=0$, extreme ones\newline

iii)${\frac{{\partial^2 V(r)}}{{\partial r^2}}}|_{extr}>0$, and stable.
\newline

\noindent being $V(r)=\left[1 +L_{T}^{2}/r^{2} - E^{2}/B\right] /A$.
Following \cite{chandra} it is found that the tangential velocity of the
test particle is

\begin{equation}
v^{tangential}=v^{\varphi }=\sqrt{\frac{rB^{\prime }}{2B}}  \label{vphi}
\end{equation}

\noindent where $^{\prime }$ means derivative with respect to $r$. It is
easy to show that if flat rotation curves are required, it arises the
following {\it \ flat curve condition} from (\ref{vphi}), that is $
B=B_{0}r^{l}$ with $l=2(v^{\varphi })^{2}$. With the {\it flat curve
condition}, metric (\ref{metric}) becomes

\begin{equation}
ds^{2}=-B_{0}r^{l}dt^{2}+A(r)dr^{2}+r^{2}d\theta ^{2}+r^{2}\sin ^{2}\theta
d\varphi ^{2}  \label{metrica}
\end{equation}
\noindent This result is not surprising. Remember that the Newtonian
potential $\psi $ is defined as $g_{00}=-exp(2\psi )=-1-2\psi -\cdot \cdot
\cdot $. On the other side, the observed rotational curve profile in the
dark matter dominated region is such that the rotational velocity $
v^{\varphi }$ of the stars is constant, the force is then given by $
F=-(v^{\varphi })^{2}/r$, which respective Newtonian potential is $\psi
=(v^{\varphi })^{2}\ln (r)$. If we now read the Newtonian potential from the
metric (\ref{metrica}), we just obtain the same result. Metric (\ref{metrica}
) is then the metric of the general relativistic version of a matter
distribution, which test particles move in constant rotational curves.
Function $A$ will be determined by the kind of substance we are supposing
the dark matter is made of. Assuming the {\it flat curve condition} in the
scalar dark matter hypothesis, we are in the position to write down the set
of field equations. Using (\ref{metrica}), the Klein Gordon equation reads 
\begin{equation}
\Phi ^{\prime \prime }+\frac{1}{2r}\left[ l+4-\frac{A^{\prime }}{A}r\right]
\Phi ^{\prime }-\frac{1}{4}A\frac{dV(\Phi )}{d\Phi }=0  \label{kg}
\end{equation}
\noindent and the Einstein equations are 
\begin{eqnarray}
\frac{A-(l+1)}{r^{2}} &=&-\kappa _{0}\left[ \frac{1}{2}\Phi ^{\prime
}{}^{2}-AV(\Phi )\right]   \label{e11} \\
\frac{1}{4r^{2}}\left[ l^{2}-\frac{A^{\prime }}{A}r\left( l+2\right) \right]
&=&-\kappa _{0}\left[ \frac{1}{2}\Phi ^{\prime }{}^{2}+AV(\Phi )\right] 
\label{e22} \\
\frac{1}{r^{2}}\left[ 1-A-\frac{A^{\prime }}{A}r\right]  &=&-\kappa _{0}
\left[ \frac{1}{2}\Phi ^{\prime }{}^{2}+AV(\Phi )\right]   \label{e00}
\end{eqnarray}

\noindent In order to solve equations (\ref{e11}-\ref{e00}), observe that
the combination of the previous equations $[(2-l)$(eq. \ref{e11})$-4$
(eq. \ref{e22})$+(2+l)$(eq. \ref{e00})$]$ implies 
\begin{equation}
V=-\frac{l}{\kappa _{0}(2-l)}\frac{1}{r^{2}}  \label{V}
\end{equation}
\noindent This is a very important result, namely the scalar potential goes
always as $1/r^{2}$ for a spherically symmetric metric with the {\it flat
curve condition}. It is remarkable that this behavior of the stress tensor
coincides with the expected behavior of the energy density of the dark
matter in a galaxy.\ We can go further and solve the field equations, the
general solution of equations (\ref{e11}-\ref{e00}) is

\[
A(r)=\left( 4-l^{2}\right) /\left( 4+C\left( 4-l^{2}\right) r^{-\left(
l+2\right) }\right) ,
\]
being $C$ an integration constant and we can thus integrate the function $
\Phi .$ Nevertheless, in this letter we consider the most simple solution of
the field equations with $C=0$. Observe that for this particular solution
the stress tensor goes like $1/r^{2}$. The energy momentum tensor is made
essentially of two parts. One is the scalar potential and the other one
contains products of the derivatives of the scalar field, both going as $
1/r^{2}$ . Furthermore, as $(\Phi _{,r})^{2}\sim 1/r^{2}$, this means that $
\Phi \sim \ln (r)$, implying that the scalar potential is exponential $V\sim
\exp (2\alpha \Phi )$ such as has been found useful for structure
formation scenarios \cite{ferre,ferre2} and scaling solutions with a
primordial scalar field in the cosmological context
\cite{ferre,ferre2,copeland,coley}
including quintessential scenarios \cite{cop2}. Thus, the particular
solution for the system (\ref{kg} -\ref{e00}) that we are considering is

\begin{eqnarray}
A &=&\frac{4-l^{2}}{4},  \label{solB} \\
\Phi  &=&\sqrt{\frac{l}{\kappa _{0}}}\ln (r)+\Phi _{0},  \label{solphi} \\
V(\Phi ) &=&-\frac{l}{2-l}\exp [{-2\sqrt{\frac{\kappa _{0}}{l}}(\Phi -\Phi
_{0})].}  \label{solv}
\end{eqnarray}

\noindent 
where (\ref{solB}) and (\ref{solphi}) approach asymptotically ($r
\rightarrow \infty$) the case
with $m=2, n=2l/(2-l)$ in the general study of the global properties of 
spherically symmetric solutions in dimensionally reduced space-times
\cite{MW}. Function $A$ corresponds to an exact solution of the Einstein
equations of a spherically symmetric space-time, in which the matter
contents is a scalar field with an exponential potential. Let us perform the
rescalling $r^{2}\rightarrow 4r^{2}/\left( 4-l^{2}\right) $. In this case
the three dimensional space corresponds to a {\it surplus of angle}
(analogous to the deficit of angle) one; the metric reads

\begin{equation}
ds^{2}=-B_{0}r^{l}dt^{2}+dr^{2}+\frac{4}{4-l^{2}}r^{2}\left[ d\theta
^{2}+\sin ^{2}\theta d\varphi ^{2}\right]  \label{main}
\end{equation}
\noindent for which the two dimensional hypersurface area is $4\pi
r^{2}\times 4/(4-l^{2})=4\pi r^{2}/(1-(v^{\varphi })^{4})$. Observe that if
the rotational velocity of the test particles were the speed of light $
v^{\varphi }\rightarrow 1$, this area would grow very fast. Nevertheless,
for a typical galaxy, the rotational velocities are $v^{\varphi }\sim
10^{-3} $ ($300km/s$), in this case the rate of the difference of this
hypersurface area and a flat one is $(v^{\varphi })^{4}/(1-(v^{\varphi
})^{4})\sim 10^{-12}$, which is too small to be measured, but sufficient to
give the right behavior of the motion of stars in a galaxy.\newline

Let us consider the components of the scalar field as those of a perfect
fluid, it is found that the components of the stress-energy tensor have the
following form 
\begin{eqnarray}
-\rho  &=&T^{0}{}_{0}=\frac{l^{2}}{(4-l^{2})}\frac{1}{\kappa _{0}r^{2}}
\label{t00} \\
P &=&T^{r}{}_{r}=\frac{l(l+4)}{(4-l^{2})}\frac{1}{\kappa _{0}r^{2}}
\label{stateeq}
\end{eqnarray}
\noindent while the angular pressures are $P_{\theta }=P_{\varphi }=-\rho $.
The analysis of an axially symmetric perfect fluid in general is given in 
\cite{nuestro}, where a similar result was found (see also
\cite{guzman1}).\newline

The effective density (\ref{t00}) depends on the velocities of the stars in
the galaxy, $-\rho =(v^{\varphi })^{4}/(1-(v^{\varphi })^{4})\times
1/(\kappa
_{0}r^{2})$ which for the typical velocities in a galaxy is $-\rho \sim
10^{-12}\times 1/(\kappa _{0}r^{2})$, while the effective radial pressure is 
$P=(v^{\varphi })^{2}((v^{\varphi })^{2}+2)/(1-(v^{\varphi })^{2})\times
1/(\kappa _{0}r^{2})\sim 10^{-6}\times 1/(\kappa _{0}r^{2})$, $i.e.$, six
orders of magnitude greater than the scalar field density. This is the
reason why it is not possible to understand a galaxy with Newtonian
dynamics. Newton theory is the limit of the Einstein theory for weak fields,
small velocities but also for small pressures (in comparison with
densities). A galaxy fulfills the first two conditions, but it has pressures
six orders of magnitude bigger than the dark matter density, which is the
dominating density in a galaxy. This effective pressure is the responsible
for the\ behavior of the flat rotation curves in the dark matter dominated
part of the galaxies. \newline

Metric (\ref{main}) is not asymptotically flat, it could not be so. An
asymptotically flat metric behaves necessarily like a Newtonian potential
provoking that the velocity profile somewhere decays, which is not the
observed case in galaxies. Nevertheless, the energy density in the halo of
the galaxy decays as 
\begin{equation}
-\rho \sim \frac{10^{-12}}{\kappa
_{0}r^{2}}=\frac{10^{-12}H_{0}^{-2}}{3r^{2}}
\rho _{crit}
\end{equation}

\noindent where $H_{0}^{-1}=\sqrt{3}/h\ 10^{6}Kpc$ is the Hubble parameter
and $
\rho _{crit}$ is the critical density of the Universe. This means that after
a relative small distance $r_{crit}\sim \sqrt{3/h^{2}}\approx 3Kpc$ the
effective density of the halo is similar as the critical density of the
Universe. One expects, of course, that the matter density around a galaxy is
smaller than the critical density \cite{suda}, say $\rho _{around}\sim
0.06\rho _{crit}$, then $r_{crit}\approx 14Kpc$. Observe also that metric 
(\ref{main}) has an almost flat three dimensional space-like hypersurface.
The difference between a flat three dimensional hypersurface area and the
three dimensional hypersurface area of metric (\ref{main}) is $\sim 10^{-12}$
, this is the reason why the space-time of a galaxies seems to be so flat.
We think that these results show that it is possible that the scalar field
could be the missing matter (the dark matter) of galaxies and maybe of the
Universe.\newline

Possibly the greatest problem with the present model is the physical
origin of the exponential potential (\ref{solv}). Firstly, its sign is
necessarily opposite to that of the exponential potentials that have been
considered in quintessence cosmologies \cite{ferre,ferre2,copeland,cop2}.
Secondly,
although exponential scalar
potentials with an overall negative sign do arise from dimensional
reduction of higher-dimensional gravity with the extra dimensions forming
a compact Einstein space of dimension $n\ge2$ \cite{MW}, such models also
constrain the parameter $l=2n/(n+2)$ to take values $l\ge1$, which are
inconsistent with its interpretation as $l=2\left(v^\varphi\right)^2$ for
velocities $v^\varphi$ of the order of magnitude of the rotation velocity
of galaxies. Similar considerations apply to the exponent of a single
exponential potential obtained by the dimensional reduction of a theory
with a higher--dimensional cosmological constant and Ricci--flat internal
space \cite{WW}. Nonetheless, exponential potentials can arise in a
variety of
ways in stringy gravity, possibly via symmetry breaking or similar
mechanisms, and so we are hopeful that a natural origin can be found for
potentials of the type considered here.\\

\acknowledgements{
We want to thank L. Arturo Ure\~na-L\'opez, Michael Reisenberger, Daniel
Sudarsky and Ulises Nucamendi for many helpful
discussions. We also want to express our acknowledgment to the relativity
group in Jena for its kind hospitality. This work is
also partially supported by CONACyT M\'{e}xico, by grants 
94890 (Guzm\'{a}n) and by the DGAPA-UNAM IN121298 (N\'{u}\~{n}ez) and by a 
cooperations grant DFG-CONACyT.}



\begin{references}
\bibitem{falta}  Peebles, P. J. E. {\it Principles of Physical Cosmology},
Princeton University Press, (1993)

\bibitem{persic}  Persic, M., Salucci, P. and Stel, F. {\it MNRAS,} {\bf
281} (1996) 27-47

\bibitem{begeman}  Begeman, K. G., Broeils, A. H. and Sanders, R. H.,
{\it MNRAS,} {\bf 249} (1991) 523

\bibitem{guzman1}  Guzm\'{a}n F. S. and Matos, T. {\it Class. Quant. Grav.,} 
{\bf 17} (2000) L9-L16. 

\bibitem{guzman2} Matos, T. and Guzm\'{a}n, F. S. {\it Ann. Phys.
(Leipzig),} {\bf 9} (2000) SI-133

\bibitem{turner}  Turner, M. S. {\it Phys. Rev}. {\bf D28}, (1983), 1243.
Ford, L. H. {\it Phys. Rev}. {\bf D35}, (1987) 2955

\bibitem{nuestro}  Matos, T., Nu\~{n}ez, D., Guzm\'{a}n, F. S. and
Ram\'{\i}rez, E., to be published. Preprint astro-ph/0005528

\bibitem{chandra}  Chandrasekhar, S. {\it Mathematical theory of black
holes,} Oxford Science Publications, (1983)

\bibitem{ferre} Ferreira, P. G. and Joyce, M., {\it Phys. Rev. Lett.,}
{\bf 79} (1997) 4740 

\bibitem{ferre2} Ferreira, P. G. and Joyce, M., {\it Phys. Rev.} {\bf
D58} (1998) 023503

\bibitem{copeland} Copeland, E. J., Liddle, A. R. and Wands, D., {\it
Phys. Rev.} {\bf D57} (1998) 4686

\bibitem{coley} Billyard, A. P. and Coley, A. A., {\it Phys. Rev.} {\bf
D61} (2000) 083503

\bibitem{cop2} Barreiro, T., Copeland, E. J. and  
Nunes, N. J. {\it Phys. Rev.} {\bf D61} (2000) 127301 

\bibitem{MW} Mignemi, S. and Wiltshire, D. L., {\it Class. Quantum Grav.},
{\bf 6} (1989) 987

\bibitem{suda}  Sudarsky, D. Private communication. To be published
elsewhere.

\bibitem{WW} Wiltshire, D. L., {\it Phys. Rev.} {\bf D44} (1991) 1100
\end{references}
\end{document}